\documentclass{article}

\usepackage{arxiv}
\usepackage[utf8]{inputenc} 
\usepackage[T1]{fontenc}    
\usepackage{hyperref}       
\usepackage{url}            
\usepackage{booktabs}       
\usepackage{amsfonts}       
\usepackage{nicefrac}       
\usepackage{microtype}      
\usepackage{lipsum}		
\usepackage{graphicx}
\usepackage{natbib}
\usepackage{doi}
\usepackage{multirow}
\usepackage{float}
\usepackage{placeins}

\title{Quantum adversarial machine learning and defense strategies:
Challenges and opportunities}

\author{
    {\hspace{1mm}Eric Yocam} \\
    The Beacom College of Computer and Cyber Sciences\\
    Dakota State University\\
    Madison, SD 57042 \\
    \texttt{eric.yocam@trojans.dsu.edu} \\
    \And
    {\hspace{1mm}Anthony Rizi} \\
    The Beacom College of Computer and Cyber Sciences\\
    Dakota State University\\
    Madison, SD 57042 \\
    \texttt{anthony.rizi@trojans.dsu.edu} \\
    \And
    {\hspace{1mm}Mahesh Kamepalli} \\
    The Beacom College of Computer and Cyber Sciences\\
    Dakota State University\\
    Madison, SD 57042 \\
    \texttt{mahesh.kamepalli@trojans.dsu.edu} \\
    \And
    {\hspace{1mm}Varghese Vaidyan} \\
    The Beacom College of Computer and Cyber Sciences\\
    Dakota State University\\
    Madison, SD 57042 \\
    \texttt{varghese.vaidyan@dsu.edu} \\
    \And
    {\hspace{1mm}Yong Wang} \\
    The Beacom College of Computer and Cyber Sciences\\
    Dakota State University\\
    Madison, SD 57042 \\
    \texttt{yong.wang@dsu.edu} \\
    \And
    {\hspace{1mm}Gurcan Comert} \\
    Engineering and Computer Science Department\\
    Benedict College\\
    Columbia, SC 29204 \\
    \texttt{Gurcan.Comert@Benedict.edu}
}

\renewcommand{\shorttitle}{Quantum Adversarial Machine Learning Defense Strategies}

\hypersetup{
pdftitle={A template for the arxiv style},
pdfsubject={q-bio.NC, q-bio.QM},
pdfauthor={David S.~Hippocampus, Elias D.~Striatum},
pdfkeywords={First keyword, Second keyword, More},
}

\begin{document}
\maketitle

\begin{abstract}
	As quantum computing continues to advance, the development of quantum-secure neural networks is crucial to prevent adversarial attacks. This paper proposes three quantum-secure design principles: (1) using post-quantum cryptography, (2) employing quantum-resistant neural network architectures, and (3) ensuring transparent and accountable development and deployment. These principles are supported by various quantum strategies, including quantum data anonymization, quantum-resistant neural networks, and quantum encryption. The paper also identifies open issues in quantum security, privacy, and trust, and recommends exploring adaptive adversarial attacks and auto adversarial attacks as future directions. The proposed design principles and recommendations provide guidance for developing quantum-secure neural networks, ensuring the integrity and reliability of machine learning models in the quantum era.
\end{abstract}


\section{Introduction}\label{sec1}
\hspace{\parindent} A defense against quantum-enabled adversarial attacks necessitates rethinking conventional machine learning security, privacy, and trust. Quantum computing fundamentally alters the classical computing threat landscape and expands the set of possible adversarial attacks. New quantum computing defense strategies will be required to protect quantum machine learning models against an evolved threat landscape \cite{1}.

\subsection{Quantum-evolved threats}
\hspace{\parindent} First, rapid advances in quantum computing hardware and algorithms. Second, several major technology companies and research labs now have functioning quantum processors with 10-100 qubits. Third, quantum systems are nearing a point for nascent quantum machine learning applications. Fourth, the field of adversarial machine learning has demonstrated that classical machine learning models have inherent vulnerabilities. Finally, quantum computers are poised to take adversarial attacks to the next level by running more sophisticated optimization algorithms to find adversarial examples.

\subsection{Quantum emergence}
\hspace{\parindent} Quantum computing is growing in relevance as the technology progresses. This technological progress is underpinned by a surge in investments from both public and private sectors, driving substantial progress in quantum hardware and algorithms. In \cite{2}, it was observed that IonQ made an 11-qubit quantum computer accessible to the public through the Amazon Braket cloud platform, and on October 1st, they revealed a 32-qubit iteration. This heightened financial backing is propelling innovation and accelerating the development of quantum technologies worldwide. 

\subsection{Quantum-driven interest}
\hspace{\parindent} The expanding community of researchers, scientists, and engineers dedicated to quantum computing is contributing significantly to this progress, bringing diverse perspectives and expertise to the field. Prior research from the authors have also shown the importance of quantum computing especially quantum machine learning (\cite{3}, \cite{4}, and \cite{5}). Alternatively, there is a dark side to quantum computing where a wide array of potential harmful applications may be created from breaking cryptography to possible unlawful drug discovery \cite{6}. 

This versatility in applications underscores the multifaceted relevance of quantum computing, positioning it as a transformative force across various industries. Additionally, the global nature of quantum computing research, characterized by active participation from countries and organizations around the world, fosters a dynamic environment of collaboration and competition. This international engagement amplifies the collective efforts towards advancing quantum technology on a global scale.

\subsection{Quantum adversarial threats}
\hspace{\parindent} Quantum adversarial machine learning (QAML) uses the principles of quantum mechanics to improve the robustness and performance of machine learning models in adversarial settings where the unique properties of quantum mechanics enable the development of new techniques for defending against adversarial attack The quantum adversarial threats can be categorized based on the capabilities of the adversary. These threats can emerge from various sources, including malevolent actors seeking to manipulate or exploit quantum algorithms and protocols. Adversarial threats are classified into three different levels based on the capabilities of the adversary summarized in Table~\ref{table1}.

\begin{table}[!ht]
\caption{Adversarial threat capability Levels.}\label{table1}
\begin{tabular}{p{0.3\linewidth}  p{0.6\linewidth}}
\toprule
\raggedright Threat Capability Level & Adversarial Behavior \\
\midrule
 Lowest & Introducing random noise or perturbations into the quantum system, Measurement attacks \\ \\
 \hline
 Intermediate & Manipulating quantum gates to alter the behavior of quantum circuits, Parameter tampering \\ \\
 \hline
 Highest & Modifying or developing quantum algorithms to subvert the learning process, Employing advanced quantum techniques like quantum crypto-analysis \\
\bottomrule
\end{tabular}
\end{table}
\FloatBarrier

At the lowest capability level, adversaries may exhibit basic disruptive behavior. This might involve introducing random noise or perturbations into the quantum system, with the aim of degrading the performance of quantum algorithms \cite{7}. Another form of low-level threat is measurement attacks, wherein adversaries tamper with the measurement process to glean information about the quantum state \cite{8}.

Moving up to an intermediate capability level, adversaries gain more sophisticated means of interference. They might attempt to manipulate quantum gates in order to alter the behavior of quantum circuits. Additionally, adversaries at this level may engage in parameter tampering, seeking to adjust the parameters of a quantum algorithm to exert influence over the outcome.

At the highest capability level, adversaries possess advanced knowledge and resources. They may delve into the design of quantum algorithms, aiming to develop or modify them in ways that subvert the learning process. Furthermore, they might employ cutting-edge quantum techniques, including quantum crypt-analysis, to break classical cryptographic schemes. This may potentially jeopardize the security of quantum machine learning systems on a broader scale \cite{9}. Beyond capability levels, an adversary's intentions and motivations is a necessary consideration summarized in Table~\ref{table2}. 

\begin{table}[!ht]
\caption{Adversarial intentions and motivation levels}\label{table2}
\begin{tabular}{p{0.3\linewidth}  p{0.6\linewidth}}
\toprule
\raggedright Intentions and Motivations & Adversarial Actions \\
\midrule
Information Gathering & Learning sensitive details about quantum state, model parameters, or training data for competitive advantage \\ \\
\hline
\raggedright Privacy Violation & Breaching privacy of training data, Inferring sensitive information about individuals or organizations \\ \\
\hline
\raggedright Infrastructure Compromise & Compromising security of quantum computing infrastructure \\
\bottomrule
\end{tabular}
\end{table}

\FloatBarrier

First, adversaries may seek to gather information, attempting to learn sensitive details about the quantum state, model parameters, or training data for competitive advantage. Second, adversaries might endeavor to manipulate the behavior of the quantum model, potentially leading to misclassifications or biased predictions. Third, privacy violation is another concern, as adversaries may attempt to breach the privacy of training data or infer sensitive information about individuals or organizations from the trained model. Finally, in the most severe cases, adversaries with substantial resources may aim to compromise the security of the entire quantum computing infrastructure.

\subsection{Quantum-evolved threat mitigation}
\hspace{\parindent}Quantum threat mitigation may be employed to counter potential adversarial threats. The potential threat mitigation may include reduce the impact of noise on quantum computations, the use of randomization, detect tampering, and the application of secure-quantum computing. For example, a mitigation to the compromise of the security for quantum computing infrastructure may be applied by continuous monitoring and auditing of the quantum computing environment for unusual activity (or signs of adversarial interference) necessary to maintain security and integrity of entire quantum computing infrastructure. As the field of quantum machine learning continues to evolve, research and best practices will be paramount the use of threat mitigation for safeguarding quantum-based systems against adversarial threats.

\subsection{Quantum landscape}
\hspace{\parindent} Quantum computing relies on qubits, which can represent both 0 and 1 simultaneously due to superposition \cite{10}. Major companies such as IBM, Google, D-Wave, IonQ, and others are actively working on quantum processors. A key metric is the number of qubits \cite{11}, \cite{12}. However, quantum computing continues to pose challenges such as high error rates, limiting the number of operations they can perform before errors accumulate \cite{13}. For example, to address high error rates, researchers are working on error correction codes, though they come with a substantial overhead in terms of qubit count \cite{14}. Various physical systems are being explored to be used for quantum computing including superconducting circuits, trapped ions, topological qubits, and photonics. Many of these systems require specialized cryogenic cooling to operate at temperatures near absolute zero \cite{15}. 

Several quantum programming languages have emerged for quantum software such as Qiskit used on IBM's systems, Cirq used on Google's systems, and Forest used on Rigetti's systems \cite{16}. Microsoft's Quantum Development Kit and other frameworks provide tools for quantum programming. Well known quantum algorithms by Shor that can be applied to factoring large numbers and by Grover that can be applied to unstructured search \cite{17}, \cite{18}. Quantum simulators enable researchers to experiment with quantum algorithms on classical computers, though they are typically limited to small-scale experiments. Additionally, companies like IBM, Google, and Rigetti offer cloud access to their quantum hardware, allowing developers and researchers to run experiments remotely.

However, the field faces several challenges. Developing effective error correction techniques is crucial for building large-scale, fault-tolerant quantum computers. Scaling up the number of qubits while maintaining qubit quality is also a significant challenge. Identifying practical applications that can leverage quantum advantage and developing efficient quantum algorithms for them is an ongoing area of research. Moreover, bridging the gap between quantum and classical systems is crucial for achieving hybrid quantum-classical computing.

\subsection{Quantum datasets}
\hspace{\parindent} Quantum computing datasets have several inherent qualities that make them well-suited for training quantum machine learning models. These inherit qualities include high-dimensional data, quantum entanglement and superposition, and quantum system simulation \cite{19}. High-dimensional data is difficult for quantum computers to represent and attack it since all of the dimensions of the data must be represented. Quantum entanglement and superposition protect datasets against introducing noise into the dataset. Quantum system simulation can be used to generate new quantum machine learning algorithms (e.g., quantum cryptography and quantum key distribution).

There is no one-size-fits-all solution for protecting quantum computing datasets. However, by curating and preparing quantum computing datasets, the quantum computing dataset can be more resilient to adversarial attacks as well as preserve the integrity of quantum training data. The protections to consider are 1) using synthetic data, 2) adding noise to data, and 3) applying post-processing of the dataset \cite{20}. First, generating artificial data (or synthetic data) can be used to create datasets that are more robust to quantum attacks. By designing synthetic data, it can be specifically tailored to resist known quantum attacks.  Second. adding noise to the data (just enough noise as to avoid degrading the dataset quality) can make it more difficult for quantum computing to break cryptographic algorithms. Finally, by applying post-processing (e.g., adversarial training) of the dataset, dataset robustness improves against quantum attacks

\subsection{Quantum algorithms}
\hspace{\parindent}
In recent years, researchers have made significant headway in developing quantum algorithms tailored for machine learning tasks. For example, quantum support vector machine (QSVM) algorithm leverages the principles of quantum computing to efficiently perform tasks like binary classification, demonstrating quantum advantage over classical SVMs algorithms under certain conditions. In addition, variational quantum algorithms have gained prominence. These algorithms employ parameterized quantum circuits and classical optimization routines. For instance, the Variational Quantum Eigensolver (VQE) is an example of a variational algorithm. It has been employed to efficiently estimate the ground state energy of molecules, a crucial task in quantum chemistry and materials science.

\subsection{Quantum challenges}
\hspace{\parindent} First, a significant challenge is the utilization of Noisy Intermediate-Scale Quantum (NISQ) devices. These quantum computers, while revolutionary, are susceptible to noise and possess a limited number of qubits. This poses challenges when implementing and training complex quantum machine learning models. For example, executing quantum error correction on NISQ devices is non-trivial and requires innovative approaches.

Second, acquiring large volumes of high-quality quantum data remains a significant hurdle. Unlike classical machine learning, which often relies on extensive datasets, quantum machine learning contends with a scarcity of quantum data. For example, in quantum chemistry applications, obtaining precise quantum measurements from experiments can be resource-intensive and technically challenging.

Third, optimizing quantum circuits for specific hardware architectures is a formidable task. For instance, when implementing quantum algorithms on different types of quantum hardware, such as superconducting qubits or trapped ions, careful circuit design and optimization are essential. This involves choosing suitable gate sets and decomposition that are compatible with the underlying hardware constraints.

Finally, interfacing quantum machine learning algorithms with classical systems and data processing pipelines represents an ongoing challenge. Achieving seamless integration between quantum and classical computing is crucial for practical applications. For example, in quantum machine learning workflows, it's vital to establish effective communication channels between classical and quantum components, ensuring smooth data processing and task execution.

\subsection{Quantum defense ethical dilemmas}
\hspace{\parindent} Quantum computing defense mechanism consideration for ethical dilemmas and concerns are a real possibility especially associated with sophisticated adversarial attacks, surveillance and oppression, and potential for autonomous weaponizing \cite{21} 

First, quantum computing defenses are designed to make machine learning models more robust to adversarial attacks. However, a possibility exists that the same hardening used with machine learning models may be also used by adversaries to construct sophisticated and effective adversarial attacks. Second, quantum computing defenses may be used by adversaries to develop new forms of surveillance and oppression that may make it increasingly difficult or potentially impossible to detect \cite{22}. Finally, quantum computing defenses may be used by adversaries to develop a capability associated with autonomous weaponizing that may make it difficult to defend against. 

\subsection{Quantum-classical adversarial attacks}
\hspace{\parindent} Quantum classifiers using machine learning, similar to quantum-classical neural networks, are susceptible to carefully crafted adversarial examples created by adding imperceptible perturbations to legitimate input data \cite{23}. Practical defense strategies, including adversarial training, can be used to mitigate the identified vulnerabilities.

Quantum-classical adversarial machine learning attacks are still evolving given the limited resources and access available for quantum computing. However, several adversarial attacks mentioned within classical machine learning studies that can also be found in quantum-classical (or hybrid) neural networks. These adversarial attacks against both classical and hybrid neural networks depending on an adversary's knowledge about the neural network. 

The adversary's knowledge levels are white-box (with full knowledge of the target classification model), black box (With no knowledge of the target classification model), and grey-box (partial knowledge of the target classification model) \cite{24}. The boundaries among white-box, grey-box, and black-box attacks are not always clear-cut (e.g. partial knowledge of a target classification model allows the adversary to launch a grey-box attack against the classification model, even if they do not have full knowledge of the classification model's architecture and parameters). The adversarial attacks include data poisoning, perturbation, membership interface, evasion, and square summarized in Table~\ref{table3}.

\begin{table}[!ht]
\caption{Adversarial attack types}\label{table3}
\begin{tabular}{p{0.3\linewidth}  p{0.6\linewidth}}
\toprule
\raggedright Attack Type & Adversary's Knowledge Level \\
\midrule
Data Poisoning & White-box \\ \\
\hline
\raggedright Perturbation & White-box, Grey-box, Black-box \\ \\
\hline
\raggedright Membership Interface & Grey-box, Black-box \\ \\
\hline
\raggedright Evasion & White-box, Grey-box, Black-box \\ \\
\hline
\raggedright Square & White-box, Grey-box, Black-box \\
\bottomrule
\end{tabular}
\end{table}

\FloatBarrier

\subsection{Quantum-classical data poisoning attack type}
\hspace{\parindent} Data Poisoning attack is achieved by corrupting the logic of the model, Manipulating the training data which is used to train the model, Injecting the data that affects the result of the model, and transfer old poisoned data to the new model \cite{25}. Data poisoning is a white-box attack because it requires the attacker to have access to the training data of the target model.

\subsection{Quantum-classical perturbation attack type}
\hspace{\parindent} Perturbation attacks occur when a single perturbation is added to the data in the deadset that misclassifies the whole dataset \cite{26}. When the perturbation is added to the model there will be a change in the learning rate, saddle point, and min and max localization. Perturbation attacks can be white-box, grey-box, or black-box attacks. In a white-box perturbation attack, the attacker has access to the target model's architecture and parameters. This allows the attacker to craft adversarial examples that are very effective at fooling the model. In a grey-box perturbation attack, the attacker has partial knowledge of the target model, such as its architecture or inputs. This makes it more difficult for the attacker to craft adversarial examples, but it is still possible. In a black-box perturbation attack, the attacker has no knowledge of the target model, other than its inputs and outputs. This makes it the most difficult type of perturbation attack to craft, but it is still possible.

\subsection{Quantum-classical inference attack Type}
\hspace{\parindent} Membership Inference attacks happen when an attacker has query access to data, and also knows the architecture of the machine learning model. After gathering some samples from a real population that matches the training data set and generating multiple input formats similar to the training data set attacker will run the model multiple times to find the type of data that behaves differently \cite{27}. Then the authors create a few shadow models that have a similar architecture as the target model and train and test these shadow models with the real data fetched from the query access and sample synthesized test data created. Membership interface attacks are grey-box or black-box attacks because they require the attacker to be able to query the target model. However, the attacker does not need to know the target model's architecture or parameters.

\subsection{Quantum-classical evasion attack type}
\hspace{\parindent} Evasion Attacks occur when the test data is manipulated with a small perturbation to the adversarial data sample. A small perturbation is added to the data samples and experimented with Multi-layer perception network on two different datasets (e.g., CICIDS 2017 and TRAbID 2017). Accuracy, Precision, Recall, and F1 scores are compared to these scores of attacked test data with the normal test data and noticed a performance drop of 21.52\% and 29.07\%. Evasion attacks can be white-box, grey-box, or black-box attacks. The type of attack depends on how much knowledge the attacker has about the target model.

\subsection{Quantum-classical square attack type}
\hspace{\parindent} Square attack is a black-box where we do not have any knowledge of the classification model. This attack model is based on the random search that iterates to find the best-optimized values for the attack\cite{28}. The square attack algorithm concept is to update the perturbations randomly for every iteration add this to the present iteration and check the improvements in the objective function. The perturbations are calculated to not cross the boundaries of the norms between l-infinity to l2. The iteration is continued until the adversarial example is found. This adversarial example is used to generate more examples of adversarial data. Square attacks can also be white-box, grey-box, or black-box attacks. The type of attack depends on how much knowledge the attacker has about the target model.

\subsection{Quantum-classical inherent vulnerabilities}
\hspace{\parindent}The inherent vulnerabilities include sensitivity to perturbations, accessibility, transferability, and data poisoning \cite{29}. A machine learning model's sensitivity to minor input alterations is also known as a lack of robustness \cite{30}. An attacker can take advantage of machine learning models when these models are openly accessible via application programming interface (API) and cloud services \cite{31}. A situation when adversarial examples can be transferred between different machine learning models represents transferability vulnerability \cite{32}. In addition, training data is vital to machine learning models so when crafted malicious data is injected into the training dataset the training data has been poisoned \cite{33}. The models that haven't done the batch normalization are vulnerable to adversarial attacks\cite{34}. 

\subsection{Quantum-classical defense}
\hspace{\parindent} Quantum-classical defense mechanisms include input transformation, randomization, and adversarial training. Input transformation defenses consist of: pixel deflection, blurring, JPEG compression, dimensionality reduction, and principal components analysis \cite{35}. The use of randomization defenses are pixel dropout, image resizing, noise injection, model ensemble, and model parameters \cite{36}. 

Adversarial examples, as quantum-classical defense mechanism, are used to explicitly train the machine learning model to improve model robustness against an attack \cite{37}. An example of the adversarial examples defense mechanism is using JPEG compression in pre-processing to neural network image classification to eliminate the noise from the dataset. JPEG compression has the ability to eliminate these distortions because adversarial attacks frequently cause perturbations that are incompatible with human psycho-visual perception. The discrete cosine transform was utilized for compression, which suppresses high-frequency data such as color hue and sharp transitions in intensity. By re-training the model with different amounts of compression on the images iteratively and with a step size of 10 until (e.g., retrained 9 times), the results showed signs of improvement. Specifically, the dataset CIFAR-10 with adversarial data was included with the FGSM attack and a perturbation of 0.02 which improved accuracy from 28.97\% to 79.57\%. As with a Deep Fool type of attack, the accuracy improved from 27.44\% to 82.71\%. The dataset GTSRB with adversarial data was included with the FGSM attack and a perturbation of 0.08 which improved accuracy from 41.00\% to 73.37\%. Again, as with the Deep Fool attack, the accuracy improved from 68.19\% to 91.70\% \cite{38}.

\subsection{Quantum-classical protections applied to quantum}
\hspace{\parindent} Several security aspects are involved with protecting Quantum-classical machine learning, including post-quantum cryptography, quantum data protection, quantum data privacy, verified quantum models, quantum malware detection, and algorithmic robustness.

\subsection{Contribution}
\hspace{\parindent} This paper surveys quantum adversarial machine learning (QAML) by demonstrating sophisticated adversarial attacks (e.g., WTC adversarial attacks) against a quantum-classical or hybrid neural network model with insights for potential defenses for QNN models, reviewing the state-of-the-art in quantum computing, quantum machine learning, and defense strategies, addressing defense strategy opportunities and challenges associated with security, privacy and trust, and recommending defense strategy, guidelines, and requirements necessary to construct quantum-secure machine learning models.

The remainder of this paper is organized as follows. First, a discussion on the characteristics of potential attacks against neural networks.  Second, a drive into an empirical study focused on attacks against the hybrid model, presenting the results and findings. Third, a focus on the quantum model, examining its characteristics and vulnerabilities to attacks. Fourth, a set of proposed quantum adversarial defense strategies, providing potential methods to mitigate risks. Fifth, the coverage of crucial aspects of security, privacy, and trust requirements in quantum models. Sixth, quantum-secure recommendations, presenting defense strategies, guidelines, and requirements associated with the security, privacy, and trust of quantum models. Finally, open issues and suggests future research, and direction for further investigation and development.

\section{Quantum-classical attacks}
\label{HybridModels}
\hspace{\parindent} In this section, a review of a quantum-classical (or hybrid) neural network model and adversarial attack characteristics that support topics covered within the quantum-classical empirical study.

\subsection{Quantum-classical model}
\hspace{\parindent} Exploration of quantum-classical neural network (HNN) or hybrid attacks not only provides insight into necessary protections but also suggests guidance for developing quantum neural network protections. A HNN model is a classical model that has been partially quantized \cite{39}. The HNN model requires a defined quantum circuit. A quantum circuit is a computational routine that consists of quantum operations on data concurrent with real-time classical computation). HNN model is simulated on conventional computer hardware rather than accessing and using costly non-conventional quantum computer hardware. HNN models also expand the attack surface area for adversaries with new capabilities that HNN models unlock. Careful design is needed to fully secure HNN models against blended classical and quantum threats such as adversarial attacks.

\subsection{White-box vs. black-box characteristic}
\hspace{\parindent} A white-box adversarial attack is when an attacker knows the HNN model’s architecture, parameters, and training dataset for an attack against the model \cite{40}. In contrast, a black-box adversarial attacks is when and attacks does not have internal HNN model knowledge. In this case, the attacker must rely on observing input-output patterns to estimate gradients or train substitutes.

\subsection{Targeted vs. untargeted characteristic}
\hspace{\parindent} A targeted attack is when a classifier within the HNN model has been targeted for misclassification as part of the adversarial attack. In contrast, an untargeted attack is when there is no target classifier within the HNN model specified as part of an adversarial attack \cite{41}. A targeted adversarial attack intends to generate false data to fool classifiers. For example, image classification uses a classifier. A classifier is a set of rules used by machine learning and deep learning models to classify data. Whereas, the untargeted adversarial attack aims to cause any kind of misclassification in the machine learning model. That is, the objective is to get the HNN model to output something incorrect.

\subsection{Distinct vs. Compounded Characteristic}
\hspace{\parindent} Distinct adversarial attacks perturb individual inputs independently to cause misclassification by the HNN model. In contrast, compounded attacks coordinate perturbations across batches of multiple inputs to bypass the HNN model's defenses that consider each sample in isolation.

\section{Classical-quantum empirical study}
\label{EmpiricalQuantum-ClassicalAttack}
\hspace{\parindent} In this section an empirical study was conducted on a classical-quantum neural network or hybrid neural network (HNN) to show that defenses against new and more powerful adversarial attacks will be effective on quantum neural network (QNN) models. Likewise, the empirical study demonstrates the need to defend against more sophisticated adversarial attacks, and by extension, defense strategies and associated design trade-offs to protect quantum neural networks. 

Effectiveness and robustness are closely related, but different, properties of machine learning models. Effectiveness refers to a model's overall performance for a particular task. In contrast, the model's robustness refers to a model's ability to perform when presented with unexpected inputs (e.g., adversarial examples). In practice, a trade-off exists between effectiveness and robustness as machine learning model properties. For example, sometimes improving accuracy leads to decreasing robustness. 

Conversely, improving robustness sometimes leads to decreasing in performance. The effectiveness of a white-box, targeted, compounded (WTC) adversarial attack against a HNN has not been studied yet. In QAML, this type of quantum-classical attack investigates a WTC adversarial attack, the effectiveness of the attack, and possibly provides insight into defenses for protecting quantum models. 

First, evaluation of white-box targeted distinct adversarial attacks evolution towards a more complex type of white-box targeted adversarial attack. In this case, the effectiveness can be measured for a combination of white-box targeted distinct adversarial attacks or WTC against an HNN model. Second, as with other technological advances, WTC attacks will require additional insight and perspective for generating new defenses with countermeasures for protecting an HNN model. Finally, the effectiveness can be determined by measuring the prediction accuracy after the WTC adversarial attack once the attack has occurred. 

The work investigates a problem with a gap expressed by looking at defense mechanisms to minimize the impact to a HNN model after a WTC adversarial attack. In past investigations, researchers have not taken into consideration the defense effectiveness after a WTC adversarial attack against a classical-quantum neural network. By investigating a defense mechanism to thwart a WTC adversarial attack, defense mechanism effectiveness after the WTC adversarial attack may lead to more sophisticated defense mechanisms with potential countermeasures for protecting a HNN model.

The empirical study contribution includes 1) the development of HNN model given that QNN model is rather limited due to the access and usage cost constraints placed on quantum computing hardware, 2) an alternative to access and usage cost constraints by means of the use of an intermediate neural network that links a CNN model to a QNN model called HNN model, 3) a HNN that is simulated on conventional computing hardware, 4) a HNN model, like CNN model, is vulnerable to a WTC adversarial attack, and 5) an effective defense mechanism can be used to minimize the impact to the HNN model after a WTC adversarial attack.

\subsection{Development environment}
\hspace{\parindent} PyTorch is a popular open-source framework and a preferred framework for deep learning and machine learning research. Cirq is also a popular open-source framework for the generic interfaces used as a simulation package for the development and execution of the HNN model. Torchattacks is a PyTorch library with a set of popular adversarial attacks used to generate adversarial examples. The Torchattacks library will be used for the WTC adversarial attack parameters for the combination of FGSM(Fast Gradient Sign Method) and PGD(Projected Gradient Descent), the combination of FGSM and CW (Carlini and Wagner's attack), and the combination of PGD and CW. FGSM attack injects small perturbation to the gradient direction of the model's loss function to generate the adversarial data\cite{42}. The PGD method runs several iterations, with each iteration's displacement in the gradient direction limiting in every single iteration. By combining both FGSM and PGD, the attack will have more ability to reduce accuracy. CW attack is similar to the PGD, which runs multiple iterations to find the best perturbation that maximizes the loss function to affect the accuracy. A combination of FGSM+CW will also affect the model more to misclassify the test data.

\subsection{Design context}
\begin{figure}[!ht]
\begin{center}
\includegraphics[width=0.5\textwidth,keepaspectratio]
{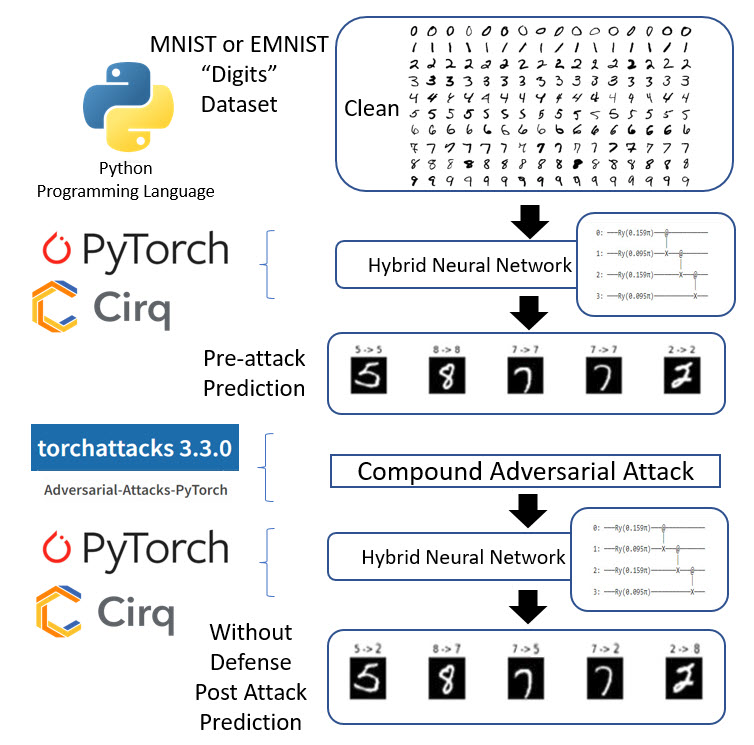}
\end{center}
\caption{Design context for the HNN model and processing without defense mechanism.}
\label{fig:figure1}
\end{figure}

\hspace{\parindent} The Python packages used for the construction and evaluation of the HNN model include PyTorch, Cirq, and Torchattacks
in Fig.~\ref{fig:figure1}. 

\begin{figure}[!ht]
\begin{center}
\includegraphics[width=0.5\textwidth,keepaspectratio]
{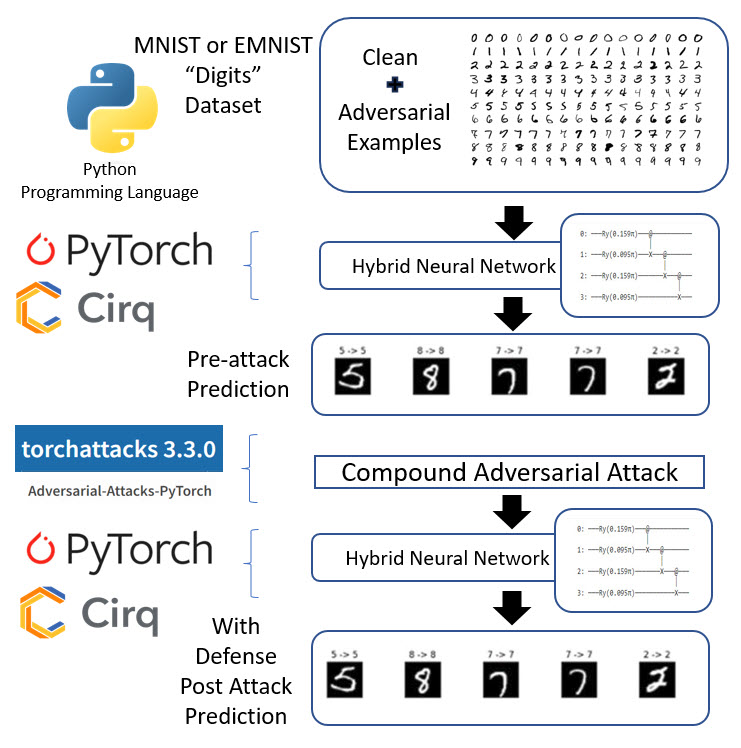}
\end{center}
\caption{Design context for the HNN model and processing with a defense mechanism.}
\label{fig:figure2}
\end{figure}

Fig.~\ref{fig:figure1} has the design of the HNN model without a defense mechanism. We can see a mismatch in the prediction of digits at the last step of Fig.~\ref{fig:figure1}. Whereas Fig.~\ref{fig:figure2} has a design context for the HNN model and processing with a defense mechanism. After using a defense mechanism, the HNN model predicted digits correctly, as illustrated in the last step of Fig.~\ref{fig:figure2}.

\subsection{HNN model}
\hspace{\parindent} The parameters for the HNN model summarized in Table~\ref{table4} includes model type, number of layers, batch size, epochs, activation function, loss function, learning rate, optimizer, digit classifier recognition, MNIST training data size, MNIST test data size, EMNIST "Digits" training data size, and EMNIST "Digits" test data size. 

\begin{table}[!ht]
\caption{Model parameters.} \label{table4}
\begin{tabular}{ | p {1 cm} | p {4 cm} | p {2 cm} | }
 \hline
 \centering No. & \centering Model or Data Parameter  &  \centering Value  \tabularnewline
 \hline\hline
 \centering 1 & \centering Model Type & \centering HNN \tabularnewline 
 \hline
 \centering 2 & \centering Number of Layers & \centering 4 \tabularnewline
 \hline
\centering 3 & \centering Batch Size  & \centering 64 \tabularnewline
 \hline
 \centering 4 & \centering Epochs  & \centering 10 \tabularnewline
 \hline
 \centering 5 & \centering Activation Functions & \centering Combination ReLU and Log Softmax \tabularnewline
  \hline
 \centering 6 & \centering Loss Function & \centering NLLoss \tabularnewline
  \hline
 \centering 7 & \centering Learning Rate  & \centering 0.001 \tabularnewline
  \hline
 \centering 8 & \centering Optimizer & \centering Adam \tabularnewline
  \hline
 \centering 9 & \centering Digit Classifier Recognizes & \centering [0,9] \tabularnewline
  \hline
 \centering 10 & \centering MNIST Training Data Size  & \centering 12,665 \tabularnewline
  \hline
 \centering 11 & \centering MNIST Test Data Size & \centering 2,115 \tabularnewline
  \hline
 \centering 12 & \centering EMNIST "Digits" Training Data Size  & \centering 48,000 \tabularnewline
  \hline
 \centering 13 & \centering EMNIST "Digits" Test Data Size  & \centering 8,000 \tabularnewline
 \hline
\bottomrule
\end{tabular}
\end{table}
\FloatBarrier

\subsection{Quantum circuit}
\begin{figure}[!ht]
\begin{center}
\includegraphics[width=0.4\textwidth,keepaspectratio]{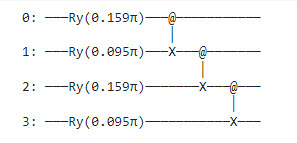}
\end{center}
\caption{The quantum circuit used for HNN model.}
\label{fig:figure3}
\end{figure}

\hspace{\parindent} A routine that computes using quantum operations on quantum data (e.g., qubits) represents a quantum circuit. The quantum circuit in Fig.~\ref{fig:figure3} is implemented using the Cirq library developed by Google. This quantum circuit consists of two components called Rotation gates and Entangling gates. Rotation gates (cirq.ry) are applied to each qubit, alternating between the angles theta and phi. Entangling gates (cirq.CNOT) are applied between pairs of adjacent qubits. To manipulate the state of individual qubits, rotation gates introduce single-qubit operations, whereas entangling gates help create a correlation between the states of different qubits. The number of qubits in the quantum circuit is determined based on the output dimension of the model. The circuit is created dynamically based on the learned parameters theta and phi.

The HNN combines the classical CNN with a quantum circuit to enhance the model's performance. The HNN takes the classical model as an input parameter and integrates it with the quantum circuit. The HNN initializes trainable parameters (theta and phi) for the quantum circuit, representing the rotation gates' angles. These parameters introduce additional non-linearity and expressiveness to the model. During the forward pass, the input data is passed through the classical CNN, and the extracted features are used as input to the quantum circuit. The quantum circuit applies quantum operations based on the learned parameters to transform the input features. The output of the quantum circuit is then processed and combined with the classical model's predictions to produce the final output.

\subsection{Datasets}
\begin{figure}[!ht]
\begin{center}
\includegraphics[width=0.4\textwidth,keepaspectratio]{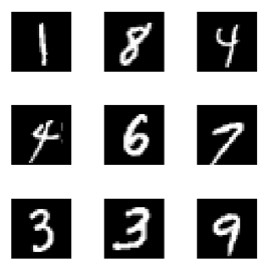}
\end{center}
\caption{The MNIST dataset sample.}
\label{fig:figure4}
\end{figure}

\hspace{\parindent} The datasets that will be used for the HNN model include MNIST and EMNIST "Digits". The MNIST dataset is popular and widely used specifically for training and testing in the field of machine learning. The MNIST dataset consists of 60,000 training examples and 10,000 examples in the test set illustrated in Fig.~\ref{fig:figure4}.  

\begin{figure}[!ht]
\begin{center}
\includegraphics[width=0.4\textwidth,keepaspectratio]{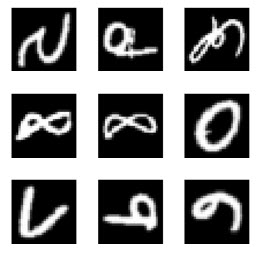}
\end{center}
\caption{The EMNIST "Digits" dataset sample.}
\label{fig:figure5}
\end{figure}

The EMNIST dataset is an extension of the MNIST dataset and consists of 240,000 training examples and 40,000 examples in the test set. Within EMNIST is a subset of the dataset called “Digits” illustrated in Fig.~\ref{fig:figure5}.  

\subsection{Procedures}
\hspace{\parindent}  
The empirical study required the development of two procedures with a set of defined steps. A 11 step procedure was created for the HNN model with no defense mechanism and a 16 step procedure was created for the HNN model with a adversarial training defense mechanism.

\subsection{Procedure for model with no defense mechanism}
\begin{figure}[!ht]
\begin{center}
\includegraphics[width=0.4\textwidth,keepaspectratio]{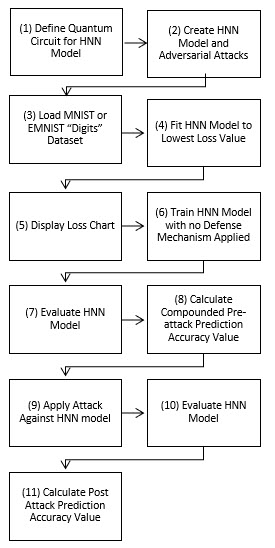}
\end{center}
\caption{Procedure for HNN model with no defense mechanism.}
\label{fig:figure6}
\end{figure}

\hspace{\parindent} A procedure was developed to take measurements of a HNN model with no defense mechanism applied. The procedure, as illustrated in Fig.~\ref{fig:figure6}, consists of eleven steps including (1) load MNIST or EMNIST "Digits" dataset, (2) define quantum circuit for HNN model, (3) create HNN model and adversarial attacks, (4) fit HNN model to lowest loss value, (5) display loss chart, (6) train HNN model with no defense mechanism applied, (7) evaluate HNN model, (8) calculate compounded pre-attack prediction accuracy value, (9), apply attack against HNN model, (10) evaluate HNN model, and (11) calculate post attack prediction accuracy value.

\subsection{Procedure for model with defense mechanism}
\begin{figure}[!ht]
\begin{center}
\includegraphics[width=0.7\textwidth,keepaspectratio]{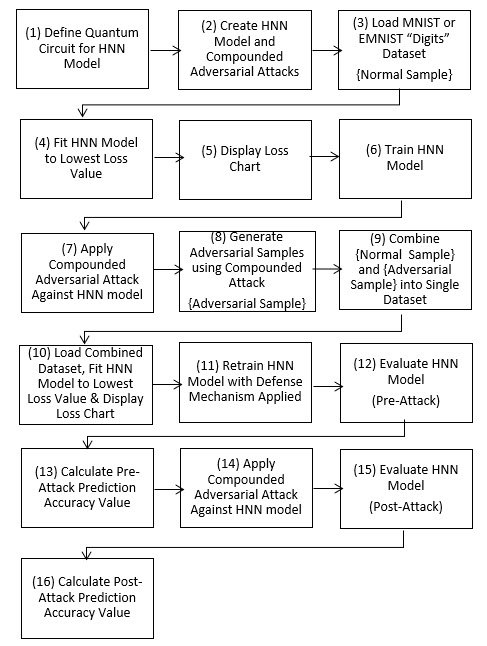}
\end{center}
\caption{Procedure for HNN model with a defense mechanism.}
\label{fig:figure7}
\end{figure}

\hspace{\parindent} A sixteen step procedure was developed, as illustrated in Fig.~\ref{fig:figure7}, for the model with defense mechanism includes (1) defining the quantum circuit for HNN model, (2) creating HNN model and compounded adversarial attacks, (3) load MNIST or EMNIST "Digits" dataset representing the normal sample, (4) fit HNN model to lowest loss value, (5) display loss chart, (6) train HNN model, (7) apply compounded adversarial attack against HNN model, (8) generate adversarial sample using compounded adversarial attack representing adversarial sample, (9) combine normal sample and adversarial sample into single dataset, (10) load combined dataset, fit HNN model to lowest loss value, and display loss chart, (11) retrain HNN model with defense mechanism applied, (12) evaluate HNN model, (13) calculate pre-attack prediction accuracy value, (14) apply compounded adversarial attack against HNN model, (15) evaluate HNN model, and (16) calculate post attack prediction accuracy value.   Adversarial training is considered to be one of the most effective defense techniques against white-box, targeted and distinct adversarial attacks. The adversarial training procedure is different from the input transformation and randomization procedures in that the adversarial training procedure focuses on combining the normal sample and the adversarial sample. The input transformation procedure focuses on the input dataset only and the randomization procedure focuses on randomizing the normal sample and randomizing the adversarial sample prior to combining with the normal sample.

\clearpage
\subsection{Empirical results}
\begin{figure*}[!ht]
\begin{center}
\includegraphics[width=0.9\textwidth,keepaspectratio]{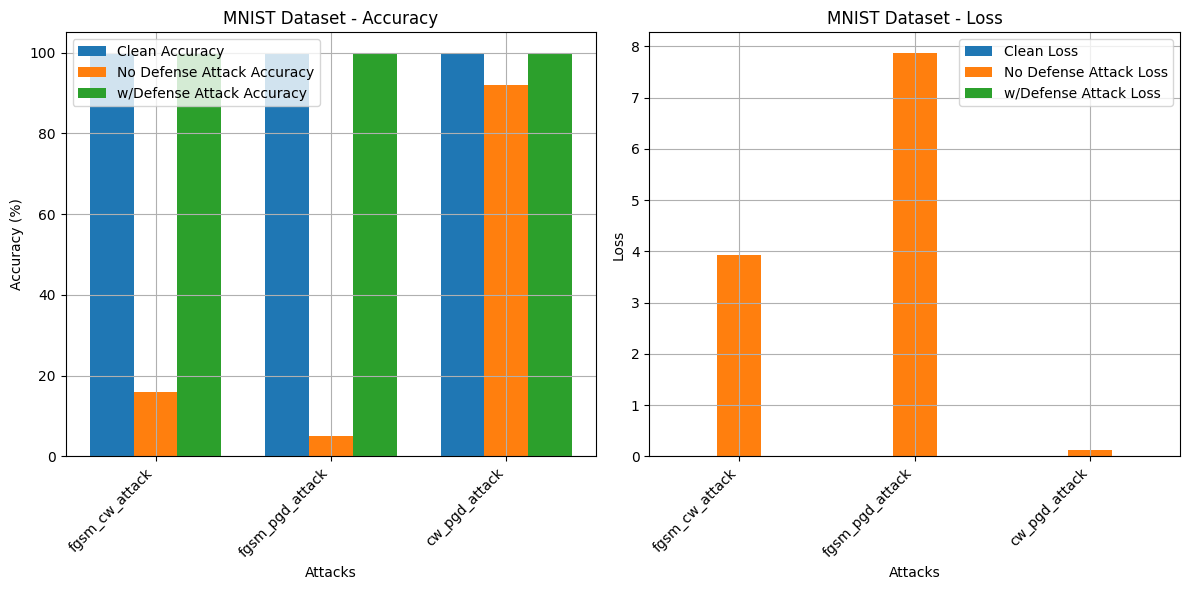}
\end{center}
\caption{MNIST dataset result comparison illustration with three compounded attacks, comparing the accuracy and loss, with and without defense techniques.}
\label{fig:figure8}
\end{figure*}

\begin{figure*}[!ht]
\begin{center}
\includegraphics[width=0.9\textwidth,keepaspectratio]{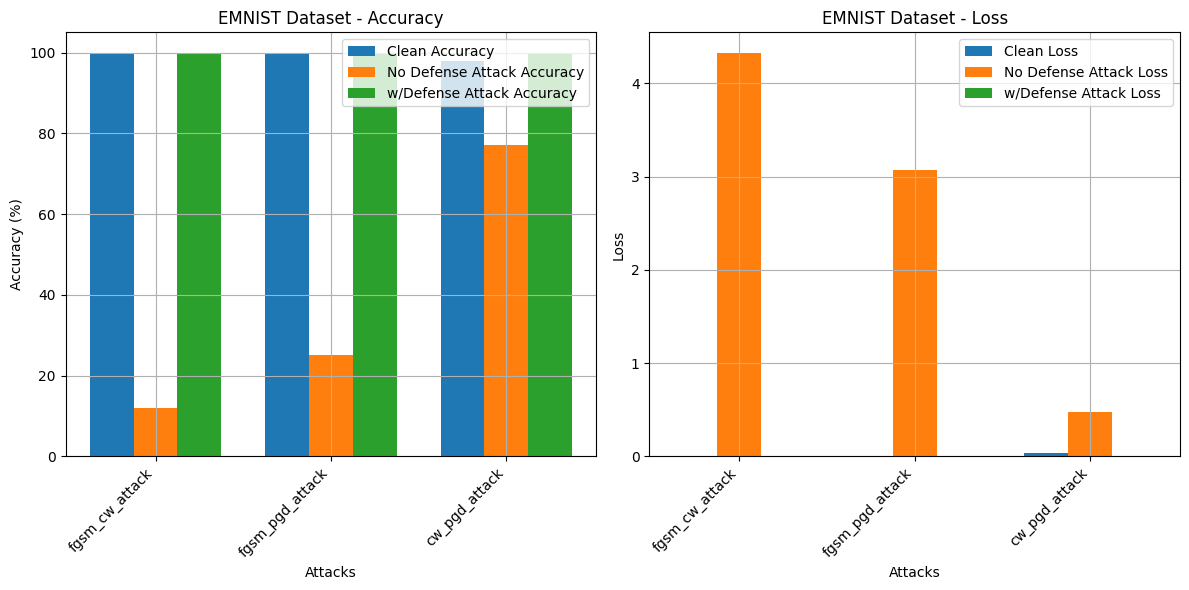}
\end{center}
\caption{EMNIST dataset result comparison illustration with three compounded attacks, comparing the accuracy and loss, with and without defense techniques.}
\label{fig:figure9}
\end{figure*}

\hspace{\parindent} Accuracy and Loss metrics(see Fig.~\ref{fig:figure8} and Fig.~\ref{fig:figure9}) from the results of HNN model for MNIST and EMNIST datasets with and without adversarial data are illustrated below. Both graphs show three attacked scenarios with three types of model techniques. A pre-attack (or Clean) accuracy indicates the dataset with no attack trained to a non-defense model, No defense attack accuracy indicates the attacked dataset is trained with a non-defense model, and defense attack accuracy indicates that attacked data is trained with a model that can resist adversarial attacks.

Table ~\ref{table5} shows the exact accuracy and loss for all the attacked dataset metrics results after training and evaluating with and without defense models. Compared to the accuracy of all attacked datasets, CW and PGD attack data have high accuracy and low loss.

We can see that the pre-attack (or clean) accuracy for the HNN model with both datasets is almost 100\%. Similarly, the loss of the HNN model is much less, nearly negligible.  In Fig.~\ref{table5}, we can compare the accuracy and loss of the HNN model attacked with and without a defense. Compared to FGSM and CW and FGSM and PGD attacks, CW and PGD attacks have less effect on the accuracy and loss for both datasets.

\begin{table*}[!ht]
    \centering
    \caption{MNIST and EMNIST dataset result in comparison with compounded attacks, comparing the accuracy and loss, with and without defense mechanisms.}\label{table5}
    \begin{tabular}{|p{2cm}||p{1.5cm}|p{1.5cm}|p{1.5cm}|p{1.5cm}|p{1.5cm}|p{2cm}|}
    \hline
    \multirow{2}{2em}{\textbf{Dataset Attack Name}} & \multicolumn{3}{|c|}{\textbf{Accuracy}} & \multicolumn{3}{|c|}{\textbf{Loss}}\\
    \hline
    & \centering Clean Accuracy & \centering No Defence Attack Accuracy & \centering With Defence Attack Accuracy & \raggedright Clean Loss & \centering No Defence Attack Loss & { \centering With Defence Attack Loss} \\
    \hline
    \centering MNIST FGSM+CW ATTACK & \centering 100\% & \centering 16\% & \centering 100\% & \centering 0.000641 & \centering 3.93 & \centering 0.00759 \tabularnewline
    \hline
    \centering MNIST FGSM+PGD ATTACK & \centering 100\% & \centering 5\% & \centering 100\% & \centering 0.0029 & \centering 7.88 & \centering 0.00848 \tabularnewline
    \hline
    \centering MNIST CW+PGD ATTACK & \centering 100\% & \centering 92\% & \centering 100\% & \centering 0.00278 & \centering 0.134 & \centering 0.0116 \tabularnewline
    \hline
    \centering EMNIST FGSM+CW ATTACK & \centering 100\% & \centering 12\% & \centering 100\% & \centering 0.00164 & \centering 4.33 & \centering 0.00218 \tabularnewline
    \hline
    \centering EMNIST FGSM+PGD ATTACK & \centering 100\% & \centering 25\% & \centering 100\% & \centering 0.000562 & \centering 3.07 & \centering 0.00315 \tabularnewline
    \hline
    \centering EMNIST CW+PGD ATTACK & \centering 98\% & \centering 77\% & \centering 100\% & \centering 0.0416 & \centering 0.478 & \centering 0.00131 \tabularnewline
    \hline
    \bottomrule
    \end{tabular}
    
\end{table*}

This compound attack will have less impact on the data because the most potent attack of these three attacks is FGSM. Meanwhile, CW and PGD have less impact attacks on the data than attacks like FGSM. Compounding one of these with FGSM will significantly impact the data. FGSM and CW leverages the efficiency of FGSM to find an initial perturbation and then refine it using CW to create a more effective adversarial attack. So, the most significant impact comes from the FGSM attack. FGSM and PGD does not provide a significant impact, as PGD already incorporates the principles of FGSM and extends it with multiple iterations. CW and PGD leverages the strengths of both attacks, such as the L2 norm minimization of CW and the iterative nature of PGD, but the attack perturbation on the data is small.

\clearpage
\section{Quantum defense strategies}
\label{QuantumModels}
\hspace{\parindent} A quantum neural network (QNN) is a parameterized quantum model executed on quantum computer hardware \cite{43}. A QNN model can be configured. However, access to quantum computer hardware and the associated computational expense of using quantum computer hardware (e.g., measured in qubits) constrains the use of QNN in practice. A qubit is a two-state (or two-level) quantum-mechanical system. A QNN adversarial attack exploits the vulnerabilities to fool QNN models using quantum algorithms and computers.

\hspace{\parindent} Several QNN defense strategies with associated opportunities and challenges are surveyed. The QNN defense strategies include quantum-assisted defense against adversarial AI, quantum generative adversarial network, variational quantum generative adversarial network, and quantum error correction.

\subsection{Quantum defense strategy trade-offs}
\hspace{\parindent} Each QNN defense strategy not only has unique strengths in leveraging quantum properties, but also challenges with integration, resource requirements, and cryptographic overheads. However, combining multiple complementary quantum defense strategies may lead to the most robust adversarial protection summarized in Table~\ref{table6}. 

\begin{table*}[!ht]
\centering
\caption{Defense strategy trade-offs}
\label{table6}
\begin{tabular}{p{0.3\linewidth} p{0.3\linewidth} p{0.3\linewidth}}
\toprule
Defense Strategy & Opportunities & Challenges \\
\midrule
\raggedright Quantum-assisted defense against AI (QDAI) & Enhance pattern recognition, detect adversarial examples, and gain insights & Requires integration with classical systems and noisy quantum processors \\
\hline
\raggedright Quantum Generative Adversarial Network (QGAN)/ Variational Quantum Generative Adversarial Network (VQGAN) & Investigate quantum generative modeling capabilities & Requires significant quantum resources, complex training, and tuning of quantum circuits \\
\hline
\raggedright Quantum Error Correction (QEC) & Enhance protections against errors/noise, stabilize quantum systems, and develop defense theory & Overhead for encoding/decoding quantum states and trade-offs associated with error correcting codes \\
\hline
\raggedright Quantum Encryption (QE) & Obscure information about QNN models and data as well as make it difficult for adversaries to interpret or exploit quantum models & Additional cryptography complexity as well as key distribution and management obstacles \\
\bottomrule
\end{tabular}
\end{table*}
\FloatBarrier

\subsection{Quantum-assisted defense strategy}
\hspace{\parindent} A quantum-assisted defense against adversarial AI (QDAI) algorithm uses a quantum computer to generate adversarial examples. The adversarial examples are used to train the machine learning model to be more robust to adversarial attacks \cite{44}. The opportunities are enhancing pattern recognition, detecting adversarial examples missed by classical models, gaining insights by using a HNN model approach. In contrast, the challenges are that this defense strategy requires integration with classical systems and noisy quantum processors may limit effectiveness.

\subsection{Quantum GAN defense strategy}
\hspace{\parindent} The quantum generative adversarial network (QuGAN) defense is a hybrid generative adversarial network (GAN) that uses a generator implemented on the quantum computer and the discriminator implemented on the classical computer \cite{45}. An similar but alternative defense to a QuGAN is the use of a variational quantum generative Adversarial Network (VQGAN) where the VQGAN is a variational GAN using a quantum circuit as generator. The opportunities include investigating these quantum generative modeling capabilities to synthesize realistic adversarial examples for training in addition to variational optimization. In contrast, the challenges are Significant quantum resources required, complex training, and tuning of quantum circuits.

\subsection{Quantum error correction defense strategy}
\hspace{\parindent} An adversarial defense is having quantum error correction (QEC) that can protect quantum information from errors. It encodes a single logical qubit into multiple physical qubits. This allows errors on individual physical qubits to be detected and correction without affecting the logical qubit \cite{46}. The opportunities are enhancing protections against errors/noise that could open attack vectors, stabilizing quantum systems, and continued development of defense theory. In contrast, the challenges are overhead for encoding/decoding quantum states and trade-offs associated with error correcting codes.

\subsection{Quantum encryption defense strategy}
\hspace{\parindent} An adversarial defense using quantum encryption leverages properties like quantum superposition and entanglement to secure communications and data. Quantum encryption makes it more difficult for adversaries to intercept, measure, or extract usable information. For example, quantum key distribution allows secure sharing of secret keys between parties to encrypt/decrypt data. The keys are encoded in quantum states that will be disturbed if improperly accessed \cite{47} The opportunities are exploring novel ways to obscure information about neural nets/data and make it more difficult for adversaries to interpret or exploit quantum models. In contrast, the challenges are additional cryptography complexity and challenges with key distribution and management.

\section{Quantum security, privacy and trust}
This section outlines a set of requirements for each design guideline pertaining to quantum security, privacy, and trust. 

\subsection{Quantum security design guidelines}
\hspace{\parindent} The quantum security design guidelines are quantum encryption, quantum compartmentalization, and quantum anomaly detection summarized in Table~\ref{table7}.

\begin{table}[!ht]
\centering
\caption{Quantum security design guidelines and requirements}
\label{table7}
\begin{tabular}{p{0.4\linewidth} p{0.5\linewidth}}
\toprule
Design Guidelines & Requirements \\
\midrule
\raggedright Quantum Encryption & Requires post-quantum crypto and quantum key distribution \\
\hline
\raggedright Quantum Compartmentalization & Requires logical segmentation, isolation, and interface minimization \\
\hline
\raggedright Quantum Anomaly Detection & Requires quantum sensors and measurement, quantum error correction, quantum data analytics, continuous monitoring, and hybrid classical quantum approaches \\
\bottomrule
\end{tabular}
\end{table}
\FloatBarrier

First, the requirements for the quantum encryption design guideline are post-quantum cryptography (PQC) and quantum key distribution (QKD). The use of PQC significantly raises the threshold for successful adversarial attacks \cite{48}. The adversary has to crack the post-quantum crypto-system before they can attempt to analyze the model or construct adversarial inputs \cite{49}. QKD enables quantum machine learning models, algorithms, and data to be robustly encrypted and authenticated, substantially raising the difficulty for adversaries attempting to steal, manipulate, or construct adversarial examples without the necessary secret keys \cite{50}. An underlying prerequisite for both PQC and QKD is having a robust random number generation (QRNG) that is a critical component for secure cryptography and key generation and necessary to encrypt quantum models and data. Quantum-resistant random number generators provide true randomness rooted in quantum properties to avoid patterns that adversaries could exploit \cite{51}.   

Second, the requirements for the quantum compartmentalization design guideline, defense-in-depth approach, are logical segmentation, isolation, and interface minimization \cite{52}. For example, critical quantum model data and parameters can be compartmentalized with only the minimal necessary access, keeping them obscured from adversaries.

Finally, the requirements for the quantum anomaly detection design guideline are advanced quantum analytics and continuous monitoring. For example, tamper detection that creates another layer of physical security monitoring for the integrity of the quantum processors themselves\cite{53}.

\subsection{Quantum privacy design guidelines}
\hspace{\parindent} Quantum privacy-preserving computing protects quantum neural networks and represents impactful defense strategies \cite{54}. At present, the existing research only considers how to protect privacy, but lacks consideration about how to obtain an optimal privacy protection strategy. The quantum privacy design guidelines are quantum data anonymization, quantum-secure multi-party computation, and quantum access controls summarized in Table~\ref{table13}.

\begin{table}[!ht]
\centering
\caption{Quantum privacy design guidelines and requirements}
\label{table13}
\begin{tabular}{p{0.4\linewidth} p{0.5\linewidth}}
\toprule
Design Guidelines & Requirements \\
\midrule
\raggedright Quantum Data Anonymization & Requires quantum scrambling and differential privacy \\
\hline
\raggedright Quantum-secure Multi-party Computation & Requires blind quantum computing and federated learning \\
\hline
\raggedright Quantum Access Controls & Require attribute-based controls and dynamic policy orchestration \\
\bottomrule
\end{tabular}
\end{table}
\FloatBarrier

First, the quantum data anonymization design guideline requires quantum scrambling and differential privacy. Differential privacy has been shown to protect user-sensitive information without significantly diminishing model accuracy \cite{55}.

Second, the quantum-secure multi-party computation requires blind quantum computing and federated learning.  Quantum secure multiparty computation refers to protocols like blind quantum computing to enable different organization to collaboratively train neural networks using confidential data in a privacy-preserving way. No input data is leaked. Quantum machine learning algorithms have built-in robustness against adversarial examples due to the quantum mechanical effects of superposition and entanglement \cite{56}.

Finally, the quantum access control requires attribute-based controls and dynamic policy orchestration.

\subsection{Quantum trust design guidelines}
\hspace{\parindent} The quantum trust design guidelines are establish hardware-based trust roots, employ trust-based protocols, and limit trust boundaries summarized in Table~\ref{table14}.

\begin{table}[!ht]
\centering
\caption{Quantum trust design guidelines and requirements}
\label{table14}
\begin{tabular}{p{0.4\linewidth} p{0.5\linewidth}}
\toprule
Design Guidelines & Requirements \\
\midrule
\raggedright Quantum Hardware-based Trust Roots & Requires quantum physical that cannot be cloned functions for hardware root of trust and tamper-resistant secure enclaves \\
\hline
\raggedright Quantum Trust-based Protocols & Requires quantum key distribution for secure channels and zero-knowledge proofs to validate operations \\
\hline
\raggedright Quantum Trust Boundaries & Requires least privilege permissions between compartments and validated communication channels \\
\bottomrule
\end{tabular}
\end{table}
\FloatBarrier

\section{Quantum-secure Recommendations}
\hspace{\parindent} In this section, an overview of related work related to quantum-secure neural networks associated with recommended quantum security, privacy and trust.

Quantum-secure neural networks are designed to be resistant to adversarial attacks. It is important to begin developing defenses against adversarial attacks on quantum neural networks now. As quantum computing continues to develop, QNN models will become more powerful and widely used, and it will be essential to ensure that they are developed using a specific set of  quantum-secure design principles.

\subsection{Quantum-secure design principle no. 1}
\hspace{\parindent}The first quantum-secure design principle is to use post-quantum cryptography summarized in Table~\ref{table10}.

\begin{table}[!ht]
\caption{Quantum-secure design principle no. 1}\label{table10}
\begin{tabular}{p{0.2\linewidth}  p{0.3\linewidth} p{0.3\linewidth}}
\toprule
\raggedright Quantum Strategy & \raggedright Quantum-secure Design Principle & Design Guideline for Quantum Security  \\
\midrule
\raggedright QDAI, QEC, QE & \raggedright Use post-quantum cryptography & Quantum-encryption, Quantum-compartmentalization, Quantum-anomaly-detection \\
\bottomrule
\end{tabular}
\end{table}
\FloatBarrier

QDAI can help to protect machine learning models from quantum attacks. QEC can help to protect quantum computers from errors. QE can help to protect data from unauthorized access. QDAI, QEC, and QE are all important for developing quantum-secure systems because they can help to protect against the unique threats that quantum computers pose.

\subsection{Quantum-secure design principle no. 2}
\hspace{\parindent}
The second quantum-secure design principle is to use quantum-resistant neural network architecture summarized in Table~\ref{table11}.

\begin{table}[!ht]
\caption{Quantum-secure design principle no. 2}\label{table11}
\begin{tabular}{p{0.2\linewidth}  p{0.3\linewidth} p{0.3\linewidth}}
\toprule
\raggedright Quantum Strategy & \raggedright Quantum-secure Design Principle & Design Guideline for Quantum Security  \\
\midrule
\raggedright QDAI, QGAN /VQGAN & \raggedright Use quantum-resistant neural network architectures & Quantum-data-anonymization, Quantum-secure-multi-part-computation, Quantum-access-controls \\
\bottomrule
\end{tabular}
\end{table}
\FloatBarrier

QDAI can be used to protect the confidentiality of the data that is used to train and deploy machine learning models. QGAN and VQGAN can be used to protect the privacy of data by generating synthetic data that is indistinguishable from real data. By using QDAI, QGAN, and VQGAN, developers can create privacy-preserving systems that can be used to protect the privacy of data

\subsection{Quantum-secure design principle no.3}
\hspace{\parindent}
The third quantum-secure design principle is for transparent and accountable development, and responsible deployment summarized in Table~\ref{table12}.

\begin{table}[!ht]
\caption{Quantum-secure design principle no. 3}\label{table12}
\begin{tabular}{p{0.2\linewidth}  p{0.3\linewidth} p{0.3\linewidth}}
\toprule
\raggedright Quantum Strategy & \raggedright Quantum-secure Design Principle & Design Guideline for Quantum Security  \\
\midrule
\raggedright QDAI, QGAN /VQGAN, QEC, QE & \raggedright Transparent and accountable development and responsible deployment &  Quantum-hardware-based-trust-roots, Quantum-trust-based-protocols, Quantum-trust-boundaries \\
\bottomrule
\end{tabular}
\end{table}
\FloatBarrier

Transparent and accountable development means that quantum systems should be developed and deployed in a way that is open and transparent, and that allows for public scrutiny and oversight.

QDAI can be used to develop machine learning models that are more trustworthy and reliable. This can be achieved by using QDAI to protect machine learning models from adversarial attacks and to improve their accuracy and robustness. QGAN and VQGAN are both powerful tools that can be used to generate synthetic data. This data can be used for a variety of purposes, such as training machine learning models and testing the robustness of security systems. Developers can help to ensure that QGAN and VQGAN are used in a transparent and accountable manner to develop secure quantum systems. QEC and QE are both important for developing quantum-secure systems. QEC can help to protect quantum computers from errors, while QE can help to protect data from unauthorized access. By using QEC and QE, developers can make quantum systems more reliable and secure, and thus reduce the risks associated with their deployment.

\section{Quantum-secure open issues and future direction}
\hspace{\parindent} In this section, several open issues have been identified and recommendation for further investigation in future work.

\subsection{Quantum-secure open issues}
\hspace{\parindent} A couple quantum security, privacy, and trust issues that remain open. First, open issues for quantum security are understanding quantum trade-offs among accuracy, complexity and security, improving secure multi-party quantum machine learning techniques, and optimizing quantum cryptography schemes for machine learning. Second, open issues for quantum privacy are developing end-to-end quantum empirical studies, and improving secure multi-party quantum machine learning techniques. Finally, open issues for quantum trust are developing quantum theories and models for robustness against adversarial attacks, and verifying integrity of quantum data and models.

\subsection{Quantum-secure future direction}
\hspace{\parindent} A recommendation for further investigation and future work consists of two potential areas to explore. Quantum-secure defenses strategies will continue to evolve for quantum adversarial machine learning. Likewise, a set of comprehensive quantum-secure defense mechanisms has proven to be a challenging task. Quantum-secure defense strategies, not covered within this paper, that are used to protect against more sophisticated types of QNN adversarial attacks include adaptive adversarial attacks and auto adversarial attacks. By exploiting the weak points of each quantum-secure defense strategies, an adversary will develop more sophisticated QNN adversarial attacks such as adaptive adversarial attack or auto adversarial attacks. As a future direction, a recommendation to explore adaptive adversarial attacks and auto adversarial attacks since these areas may be worth broadening the investigation into quantum-secure defenses mechanisms.

\section{Conclusion}
\hspace{\parindent} This paper aims to catalyze multi-disciplinary research in the increasingly crucial field of quantum adversarial machine learning. The findings from the empirical study suggest that a QNN model is just as vulnerable to adversarial attacks as an HNN model. The insights gained from the study, particularly the demonstration of sophisticated adversarial attacks on HNN models, emphasize the urgent need for developing robust defense strategies for QNN models.

To address this challenge, a set of defense strategies, including QDAI, QGAN, QEC, and quantum encryption, were proposed. The trade-offs between opportunities and challenges for each defense strategy were thoroughly analyzed and summarized. These recommended defenses for quantum adversarial machine learning provide valuable guidance for both near-term priorities and long-term investigations in this rapidly evolving field.

The technical contributions of this paper are three-fold. First, it presents an empirical study comparing the vulnerabilities of QNN and HNN models to adversarial attacks. Second, it proposes a comprehensive set of defense strategies tailored to the unique challenges posed by quantum adversarial machine learning. Third, it provides a detailed analysis of the trade-offs associated with each defense strategy, enabling researchers and practitioners to make informed decisions when implementing these defenses.

As the field of quantum computing continues to advance, the development of robust and effective defense strategies against adversarial attacks on quantum machine learning models becomes increasingly critical. This paper lays the groundwork for future research in this area, offering valuable insights and recommendations for the development of quantum-secure machine learning systems.

\bibliographystyle{apalike}
\bibliography{article}  

\end{document}